\documentclass[aps,prb,twocolumn,unsortedaddress]{revtex4}
\usepackage{graphicx}% Include figure files
\usepackage{dcolumn}
\begin{document}

\title{Acoustic vibrations of a silica-embedded gold nanoparticle:
       elastic anisotropy}

\author{Daniel B. Murray}
\email{dmurray@ouc.bc.ca}
\affiliation{Department of Physics, Okanagan University College,
             Kelowna, British Columbia, Canada V1V 1V7}
\author{S. Dhara}
\email{dhara@igcar.ernet.in}
\author{T. R. Ravindran} 
\author{K. G. M. Nair}
\author{S. Kalavathi}
\affiliation{Materials Science Division,
             Indira Gandhi Centre for Atomic Research,
             Kalpakkam 603 102, India}
\author{Lucien Saviot}
\email{Lucien.Saviot@u-bourgogne.fr}
\affiliation{Laboratoire de R\'eactivit\'e des Solides,
             UMR 5613 CNRS - Universit\'e de Bourgogne\\
             9 avenue A. Savary, BP 47870 - 21078 Dijon - France}
\author{Eug\`ene Duval}
\affiliation{Laboratoire de Physico-Chimie des Mat\'eriaux
Luminescents, CNRS and Universit\'e Lyon I, B\^atiment 203,
43 Boulevard du 11 Novembre 1918, 69622 Villeurbanne Cedex, France}

\date{\today}

\begin{abstract}
Spherical gold nanoclusters are grown by 1.8~MeV~Au$^{++}$
ion implantation into an amorphous silica matrix and
subsequent air annealing at 873~K.  Ultraviolet and visible
light absorption confirms the presence of a dipolar surface
plasmon peak at the expected location for a gold sphere in
silica.  Grazing incidence X-ray diffraction shows peaks
corresponding to fcc gold with lattice spacing close
to that of bulk gold, as well as allowing estimation of
cluster size using Scherrer's formula.  Low frequency Raman
scattering reveals a relatively narrow peak, suggesting a
narrow distribution of nanocrystal diameters.  Acoustic
phonon frequencies corresponding to the spheroidal
quadrupolar vibrations of a continuum sphere with the
anisotropic elasticity of gold are calculated using a novel
method of molecular dynamics and extrapolation to the
continuum limit using relatively small numbers of point
masses.  The study confirms high energy ion implantation as
a method capable of producing gold nanocrystals with
spherical shape and well controlled size.
\end{abstract}


\maketitle

\section{Introduction}

A free elastic sphere has vibrational modes whose lowest
frequency is on the order of the speed of sound divided
by the sphere radius.  This classic problem of elastic
mechanics was solved by Lamb in 1882 \cite{Lamb1882} for
the case of a homogeneous sphere with isotropic
elasticity.

Ninety years later, this solution found application in
explaining anomalies of the specific heat of powders of
spherical lead particles 2 to 4~nm in size.\cite{Novotny72}
Direct observations of vibrational modes of 
mixed spinel
(MgCr$_2$O$_4$-MgAl$_2$O$_4$)
microcrystallites 22 to 35~nm in
diameter\cite{Duval86} became possible by shining an intense
laser beam onto the sample and carefully analyzing the
scattered light to detect scattered frequencies slightly
above and slightly below the laser frequency.  The frequency
shift of the light is equal to a vibrational frequency of
the spinel microcrystallite.  Since then, Raman and
Brillouin scattering have remained powerful tools in the
characterization of ``nanoparticles'' -- particles with
diameters on the order of a few to a few dozen nanometers.

The nanoparticles in these experiments are often not free,
but rather embedded in solid materials such as glass.  The
Lamb solution does not directly apply since it assumes zero
traction force boundary conditions of the sphere
(\textit{i.e.} a free surface).  Initial
attempts\cite{Tamura82,Ovsyuk96} to generalize the
calculation to an isotropic elastic continuum sphere
embedded in an infinite elastic matrix are in
error.\cite{Murray04}  But an earlier
calculation\cite{Dubrovskiy81} in the geophysics literature
had correctly solved the problem and showed that
embedded sphere vibrations can be acurately approximated
with complex-valued pseudo-frequencies $\omega$ where
Re$(\omega)$ is the center frequency of the Raman peak
and 2Im$(\omega)$ is the full width
at half maximum (FWHM).  Correct application of this
solution to nanoparticles was made only
recently.\cite{DelFatti99,Verma99,Saviot04}

Vibrations of an embedded sphere are described in terms of
a displacement field $\vec{u}(\vec{r},t)$ which equals
$\vec{r}-\vec{R}$ where $\vec{R}$ is the equilibrium position
of a material point and $\vec{r}$ is its displaced position.
Due to the spherical symmetry, it is convenient to choose the
origin at the center of the sphere and use spherical
coordinates $r$, $\theta$ and $\phi$.  The time dependence
is $\mathrm{exp}(-\mathrm{i} \omega t)$ where $\omega$ is a
complex number.  In general, a vector field $\vec{u}$ is the
sum of a zero-curl field $\vec{\nabla} \Phi$ and a zero
divergence field $\vec{\nabla} \times \vec{\Xi}$ where
$\Phi(\vec{r})$ is a scalar field and $\vec{\Xi}(\vec{r})$
is a vector field.

In the situation of an isotropic, homogeneous elastic medium
with transverse speed of sound $v_T$ and longitudinal speed
of sound $v_L$, normal modes are found by solving two
uncoupled equations
\begin{equation}
( \nabla^{2} + k_{L}^{2} ) \Phi = 0
\label{A4}
\end{equation}
and
\begin{equation}
( \nabla^{2} + k_{T}^{2} ) \vec{\Xi} = \vec{0}
\label{A5}
\end{equation}
where $k_L$ = $\omega / v_L$ and $k_T$ = $\omega / v_T$.

For convenience, the original Lamb solution will be called
the Free Sphere Model (FSM).  The isotropic sphere embedded
in an infinite matrix will be called the Complex Frequency
Model (CFM).

Whether for FSM or CFM, there are some solutions of the
equation of motion with $\vec{\nabla} \cdot \vec{u} = 0$.
Lamb called these the ``first class'' and the modern word is
``torsional'' (TOR).  Apart from torsional modes, an
important special case are radial or ``breathing'' modes, in
which $\vec{u}$ is purely along the $r$-direction and
depends only on $r$ and not on $\theta$ or $\phi$.  In this
case $\vec{u}$ is of the form $\vec{\nabla} \Phi$.  The
remaining modes of what Lamb called the ``second class'', 
now called ``spheroidal'' (SPH), have a more complicated form:
\begin{equation}
\vec{u} = \vec{\nabla} \Phi + \vec{\nabla} \times \vec{\nabla} \times (\vec{r} \psi)
\label{A7}
\end{equation}
where $\Phi$ and $\psi$ are scalar fields.
Apart from the breathing modes, spheroidal modes are neither
zero curl nor zero divergence.

It can be said that torsional modes are purely transverse
in nature.  As a result their frequencies depend only on
$v_T$.  The breathing modes are purely longitudinal, but
mode frequencies depend on both $v_T$ and $v_L$.  The
remaining spheroidal modes are a mixture of motions, but
in most cases their frequencies depend primarily on $v_T$
and only weakly on $v_L$, so they are predominantly
transverse in character.  For a generic material with
Poisson ratio $\frac{1}{3}$, 80\% of all modes, counting
both TOR and SPH, are more transverse, while 20\% are more
longitudinal. 

To systematically label FSM or CFM modes, it is convenient
to use index $q \in $\{TOR,SPH\}.  In addition, modes have
an angular momentum number $\ell \in \{0, 1, 2... \}$ and
a $z$-angular momentum $m$ where $-\ell \leq m \leq \ell$.
For brevity we consider only the $m = 0$ case below.
But it is important to keep in mind that FSM and
CFM modes have degeneracy $2 \ell + 1$.
There is also a solution index $n \in \{0, 1, 2...\}$.  The
scalar field $\Phi$ satisfying Eq.~(\ref{A4}) has the form
\begin{equation}
\Phi(r,\theta) = j_{\ell}(k_L r) P_{\ell}( \cos \theta )
\end{equation}
where $j_\ell$ are spherical Bessel functions and
$P_\ell$ are Legendre polynomials.  The scalar
field $\psi$ in Eq.~(\ref{A7}) has the form
\begin{equation}
\psi(r,\theta) = j_{\ell}(k_T r) P_{\ell}( \cos \theta ).
\end{equation}

A given normal mode, whether FSM or CFM, will be denoted
($q$,$\ell$,$m$,$n$) and its frequency is
$\omega_{q \ell n}^{\mathrm{FSM}}$ or
$\omega_{q \ell n}^{\mathrm{CFM}}$.

The first applications of the correct CFM solution for an
isotropic continuum sphere embedded in a matrix were
restricted to spheroidal modes.\cite{DelFatti99,Verma99}
Group theoretical arguments\cite{Duval92} show that in the
absence of any kind of anisotropy or non-sphericity, and in
the limit that the nanoparticle is much smaller than the
wavelength of the laser light, only the modes (SPH,$\ell$=0)
and (SPH,$\ell$=2) can lead to Raman scattering.  Correct
calculations of the CFM torsional and spheroidal
$\ell \neq 0$ mode frequencies were reported
later.\cite{Saviot04}  Even so, the usage of complex-valued
frequencies of normal modes is problematic.
There is no corresponding displacement field.

Clarification of the significance of CFM calculations was
obtained through an exhaustive calculation of all of the
real-valued frequencies of an isotropic spherical
nanoparticle embedded in a large, but finite matrix with
a free spherical outer surface of radius
$R_m$.\cite{Murray04}  This is called the Core-Shell Model
(CSM).  Taking the limit $R_m \rightarrow \infty$, the
number of CSM modes in a given range of frequency becomes
infinite, and has the Debye density of states.  However,
assuming uniform random phase excitation of the CSM modes,
the mean squared displacement within the nanoparticle
exhibits peaks whose positions agree closely with the real
part of CFM modes, and whose FWHM agree closely with twice
the imaginary parts of CFM modes.  Thus, CFM calculations
have a solid basis and utility, even if the corresponding
CFM modes must be treated with caution.

Both FSM and CFM have limitations of applicability to real
systems.  (1) Nanoparticles are not truly spherical in
shape, although the dynamics of most elaboration processes
make them approximately that way and transmission electron
microscopy (TEM) images confirm this. (2) For small
nanoparticles (smaller than those in the present study),
the wavelength of even the low-lying modes
is not long compared to interatomic spacing, so that we are
not close to the center of the Brillouin zone and the
acoustic limit. (3) There will be deviations of the
properties of the nanoparticle material from the same
material in the bulk due to surface effects. (4) In a given
sample, the radii of nanoparticles will have a distribution,
rather than a single value. (5) The interface between the
nanoparticle and the matrix is not always correctly
described in terms of continuity of the displacement field
and components of the stress tensor.\cite{Bettenhausen03}
(6) The speed of sound within the nanoparticle depends on
the direction of propagation if the elasticity of the
material is anisotropic.  Only the last issue is addressed
in this paper.
          
\section{Elastic Anisotropy}

Since these words are frequently misused, we recall that
``isotropic'' means ``unchanging under rotation'' while
``homogeneous'' means ``unchanging under translation''.
Although bulk polycrystalline gold is elastically isotropic,
a single crystal of gold is not.  In fact, the noble metals
(Au, Ag, Cu) are remarkably anisotropic.  Depending on the
direction of propagation in the crystal, a transverse
acoustic (low frequency) sound wave can travel at speeds
varying from 867~m/s to 1482~m/s. The longitudinal speed of
sound varies less, ranging from 3147~m/s to 3440~m/s.  But
since all FSM and CFM modes depend on $v_T$, it must be
expected that the strong transverse anisotropy will play an
important role.

Application of FSM to elastically anisotropic crystals (in
fact, no crystalline material is elastically isotropic) goes
back many years.  Since the theory asked for $v_T$ and
$v_L$, it was common to simply use the speeds of sound in
the bulk polycrystalline material, sometimes with
apology\cite{Fujii91} but usually without.  One approach is
to make use of the speeds of sound along the principal
crystal directions, (\textit{e.g.} $<$100$>$, $<$110$>$
and $<$111$>$ in silicon\cite{Fujii96}) so as to get several
different estimates of the frequencies.

The elastic mechanical problem of calculating the
vibrational modes of a free sphere with anisotropic
elasticity can be set up analytically for certain special
kinds of anisotropy,\cite{Chen00} but not for cubic
crystalline elasticity such as that of fcc gold.  Numerical
techniques for finding eigenfrequencies of an object with
arbitrary elasticity and a variety of special shapes have
long been known.\cite{Demarest71,Visscher91}

Our numerical method used here varies only slightly from
a previous one.\cite{Saviot04}  Its features are: (1)
arbitrary cubic crystalline elasticity including arbitrarily
high anisotropy (unlike the earlier method\cite{Saviot04}
which numerically blows up when faced with the anisotropy
of a noble metal, but worked for silicon); (2) ease of
handling a nanoparticle of arbitrary shape; (3) ready
extendability to inhomogeneous density and elasticity;
(4) generalizability (with simple modification) to
hexagonal and other crystal symmetries; (4) self-diagnosis
of numerical convergence; (5) rapid convergence;
(6) conceptual simplicity.

Our approach can be described as a Finite Difference Time
Domain (FDTD) calculation, and also as molecular dynamics
(MD), the former phrase being common in the engineering
literature.  The continuum sphere of radius $R_p$ and
density $\rho_p$ is approximated by $N$ equal point masses
arranged in a simple cubic (SC) lattice.  The point masses
do not represent atoms.  For calculations reported
here, $N$ was as high as 8000.  The use of an SC lattice is
associated with the cubic symmetry of the elasticity but
does not preclude simulations of objects of crystalline
material with fcc, bcc or diamond structures, as well as
isotropic solids such as glass.  The method as detailed here
applies only to the homogeneous cubic case.  The method is
unrelated to calculations which model the microscopic
inter-atomic forces such as the Tersoff potential for
silicon or the Keating model.  Rather, it is a
continuum model of the material, involving only the bulk
elastic constants and the density.

\section{Molecular Dynamics Method}

Our method is a molecular dynamics calculation.
While this approach can readily be extended to the
more general case of an object of arbitrary shape
with inhomogeneous anisotropic elasticity, we choose for
brevity only to present the case of a spherical
object with cubic crystalline elasticity.
A simple cubic lattice with lattice parameter $a$ of $N$
identical point particles of mass $\rho a^3$
interacting through ``springs'' has particle positions
$\vec r_j$ integrated in time and Fourier transformed in
order to obtain mode frequencies.  Each particle has six
first neighbors, twelve second neighbours, eight third
neighbors and so on.  Every pair
$(i,j)$ of first neighbors are coupled by potential energy 
$\frac{1}{2} k_{sp1} (\| \vec r_i - \vec r_j \|-a)^2$.
Second neighbor pairs $(i,j)$ are coupled by
$\frac{1}{2} k_{sp2} (\| \vec r_i - \vec r_j \|-\sqrt{2} a)^2$.
Third neighbor pairs $(i,j)$ are coupled by
$\frac{1}{2} k_{sp3} (\| \vec r_i - \vec r_j \|-\sqrt{3} a)^2$.
Fourth neighbor pairs $(i,j)$ are coupled by
$\frac{1}{2} k_{sp4} (\| \vec r_i - \vec r_j \|-\sqrt{4} a)^2$.
Fifth neighbor pairs $(i,j)$ are coupled by
$\frac{1}{2} k_{sp5} (\| \vec r_i - \vec r_j \|-\sqrt{5} a)^2$.
Pairs are not double counted.
Cubic octets are coupled by
$ ( k_{8pt} / 4 a^2 ) (D - 6 a^2)^2$
where
$ D = \sum_{j=1}^{8} \| \vec R_{cm} - \vec r_j \|^2 $
and
$ \vec R_{cm} = \frac{1}{8} \sum_{j=1}^{8} \vec r_j $.

No microscopic physical interpretation is offered
for the $k_{8pt}$ term.  Its purpose it to allow reproduction
of the bulk elastic constants rather than to represent the
potentials due to atomic bonding among atoms.  It is possible
to approximate interatomic interactions among atoms in certain
special kinds of crystals using model
potentials.\cite{Keating66,Stillinger85,Tersoff89}  Our
approach is not limited to certain kinds of crystals, but
rather to continuum materials with completely general cubic
elasticity.  For this purpose, it is numerically more
efficient to use the potentials that we employ.

Although we have introduced six force constants,
($k_{sp1}$, $k_{sp2}$, $k_{sp3}$, $k_{sp4}$, $k_{sp5}$,
and $k_{8pt}$) only $k_{sp1}$, $k_{sp3}$,
and $k_{8pt}$ are used for the numerical calculations of
this paper.  A previous paper\cite{Saviot04} used
$k_{sp1}$, $k_{sp2}$, and $k_{8pt}$, but was limited to
materials of lower elastic anisotropy as will be explained
below.  It is advantageous to use the earlier method for
less isotropic materials since numerical convergence is
faster as $N$ increases.

The next step is to show how the force constants are
related to the bulk elastic constants of the material
of the nanosphere.  A cubic crystal (in particular,
any material with bcc, fcc, diamond or zincblende
crystal structure) has three elastic constants:
$C_{11}$, $C_{12}$, and $C_{44}$.  For an isotropic material,
$C_{12} = C_{11} - 2 C_{44}$.  The degree of elastic
anisotropy is quantified by the Zener anisotropy ratio,
$A_Z$, where \begin{equation}
A_Z = \frac{2 C_{44}}{C_{11}-C_{12}} 
\end{equation}
Elastic isotropy corresponds to $A_Z = 1$, and anisotropic
materials can have $A_Z$ above or below 1.  The Zener
anisotropy of gold is 2.92.  Silver and copper have
comparable anisotropies.

The bulk modulus of an arbitrary cubic elastic material is
$B = ( C_{11}+2 C_{12} ) / 3$.  Let $V$ denote volume.  A
small isotropic dilatation $ \delta = dV/V $ stretches all
neighbor distances proportionally and leads to stored energy
$ \frac{1}{2} V B \delta^2 $, so that we obtain
\begin{equation}
\label{eqB}
a (C_{11} + 2 C_{12} ) = k_{sp1} + 4 k_{sp2} + 4 k_{sp3} + 4 k_{sp4} + 20 k_{sp5} + 24 k_{8pt}
\end{equation}

Considering that a longitudinal plane wave moves along the
$x$-axis at speed $ \sqrt{C_{11} / \rho} $, we find
\begin{equation}
\label{eqC11}
a C_{11} = k_{sp1} + 2 k_{sp2} + \frac{4}{3} k_{sp3} + 4 k_{sp4} + 13.6 k_{sp5} + 8 k_{8pt}
\end{equation}

Finally, considering that a transverse plane wave moves
along the $x$-axis at speed $ \sqrt{C_{44} / \rho} $, we
find
\begin{equation}
\label{eqC44}
a C_{44} = k_{sp2} + \frac{4}{3} k_{sp3} + 3.2 k_{sp5}
\end{equation}

By subtracting Eq.~\ref{eqC11} from Eq.~\ref{eqB}, we obtain
\begin{equation}
a C_{12} = k_{sp2} + \frac{4}{3} k_{sp3} + 3.2 k_{sp5} + 8 k_{8pt}
\end{equation}

Note that if $k_{8pt} = 0$, $C_{12}$ and $C_{44}$ are exactly
the same.  This has been numerically verified up to tenth
neighbours.  It is coincidentally the case that $C_{12}$ and
$C_{44}$ are almost the same for NaCl,  
so the elastic properties
of NaCl could be modelled without the $k_{8pt}$ term.  
In addition,
$C_{12}$ and $C_{44}$ are equal for isotropic materials with
Poisson ratio $\frac{1}{4}$, such as polycrystalline Cerium,
once again allowing modelling with two-body forces alone.

For other materials, it is essential to include an
interaction apart from two-particle forces.  This is the
reason for including the $k_{8pt}$ potential term.  Eight
masses must be involved in order to have a term that in
itself respects cubic symmetry.  However, it would also be
workable though not obviously advantageous to have used a
three body interaction term.

Three interaction terms, including $k_{8pt}$, must be used
to exactly fit the three elastic parameters
$C_{11}$, $C_{12}$, and $C_{44}$.
If, as in earlier work\cite{Saviot04}, 
only $k_{sp1}$, $k_{sp2}$, and $k_{8pt}$ are used, then
\begin{equation}
k_{sp1} = a ( C_{11} - ( C_{12} + C_{44} ) ),
\end{equation}
Note that materials with Zener anisotropy above 2 will have
negative $k_{sp1}$.  This leads to instability of molecular
dynamics simulations of the lattice.

Another option is to use only $k_{sp1}$, $k_{sp3}$, and
$k_{8pt}$.  Inversion allows the three force constants to be
obtained for a general cubically elastic material:
\begin{equation}
k_{sp1} = a ( C_{11} - C_{12} ),
\end{equation}
\begin{equation}
k_{sp3} = \frac{3a}{4} C_{44} 
\end{equation}
and
\begin{equation}
k_{8pt} = \frac{1}{8} a ( C_{12} - C_{44} ).
\end{equation}
Note that $k_{sp1}$ and $k_{sp3}$ are never negative.
It can be the case that $k_{8pt}$ is negative for some
materials, but in practice this does not lead to instability
of the molecular dynamics simulations.  As a result, any
cubic material can be simulated.

The frequency dependence of a given mode is $f(N)$.
A sequence of simulations with
$N$ varying from a few hundred to 8000 is used to
extrapolate $f(\infty)$, which is the continuum limit.
Accurate results are obtainable with just a few thousand.

The correctness and degree of convergence
of our MD method was checked by comparing its results to
those of the Lamb solution for isotropic elastic materials
of various Poisson ratios.  The results of the two methods
agree closely.

\section{Molecular Dynamics Results}

A sequence of simulations with increasing $N$ is performed
in order to extrapolate the limit of infinite $N$.  The
purpose of the calculation is to solve for the vibrational
frequencies of a continuum elastic sphere, and does not
take into account the crystal structure of the material or
dispersion (frequency dependence of sound speed) of phonons.

The time duration of the simulation is determined by the
frequency resolution that is desired.  The initial velocity
distribution corresponds to an FSM mode.  The $z$-axis of
the desired FSM mode is randomly oriented for each run
relative to the axis of the cubic lattice.

A Fourier transform is carried out of the motion of several 
randomly selected mass points along various axes.  A plot
is then made of the power spectrum versus $1/N$.  This
permits extrapolation of the frequency of each mode to
infinite $N$.  The extrapolation is done visually.

The elastic constants used \cite{LBAu} for monocrystalline
fcc gold at 300~K were
$C_{11}$~=~191~GPa, $C_{12}$~=~162~GPa, $C_{44}$~=~42.4~GPa,
and $\rho$~=~19.283~g/cc.

Table~\ref{TAB1} shows results of the molecular dynamics
simulations.  Frequencies are presented in terms of the
ratio $\omega R$.  For an isotropic material, it is
beneficial to use dimensionless frequencies such as
$\eta = \omega R / v_T$ and $\xi = \omega R / v_L$.
In this problem, the anisotropy of the elasticity means
that there is no convenient sound speed scale to form
a dimensionless frequency.  However, expressing results
in terms of $\omega R$ exploits the scale invariance of
the problem as long as continuum elasticity is assumed.

The frequencies in Tab.~\ref{TAB1} are classified
according to the symmetry of the initial disturbance used
to excite them.  In each case, velocity fields of
spheroidal FSM modes of angular momentum
$\ell$ are used as the initial disturbance.  The $z$-axis
of the velocity field is randomized with respect to the
axes of the lattice for each run.
The mode degeneracies in Tab.~\ref{TAB1} were found
by gradually distorting the shape of the sphere so as to
break all symmetry, and noting into how many frequencies
each mode splits.
The $\ell$=0 spheroidal mode frequency is calculated, but the
Raman scattering strength is typically much smaller than for
the $\ell$=2 spheroidal mode.

\begin{table}
\caption{\label{TAB1} Frequencies of selected vibrational
modes of an elastic continuum sphere with the elastic
constants of single crystal gold of radius $R$ are shown.
The value given is $\omega R$ (in m/s) where
$\omega$ is the frequency in radians per second and $R$
is the nanosphere radius in metres.}
\begin{tabular}{|c|c|c|c|c|c|}
\hline
$q$  & $\ell$ & $n$ & $\omega R$ & degeneracy & symmetry\\
\hline
SPH  & 2 & 0 & 2342 & 3 & $T_{2g}$ \\
\hline
SPH  & 2 & 0 & 3779 & 2 & $E_{g}$  \\
\hline
SPH  & 0 & 0 & 10315 & 1 & $A_{1g}$ \\
\hline
\end{tabular}
\end{table}

\begin{figure} 
\includegraphics[width=\columnwidth]{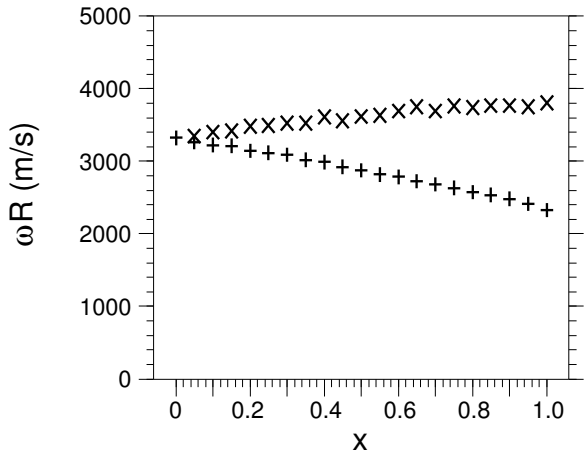}
\caption{\label{FIG1} 
Perturbed frequencies of the initially fivefold degenerate
quadrupolar spheroidal modes of a free sphere are shown
as the parameter $x$ is varied.  $x$ = 0 corresponds to
directionally averaged isotropic gold while $x$ = 1
corresponds to true anisotropic gold.  The upper
and lower branches are twofold and threefold degenerate respectively.}
\end{figure}

It would be expected on the basic of group theoretical
considerations that a 5-fold degenerate (SPH,2) mode
of an isotropic sphere would be broken into two different
levels by lowering the elastic symmetry to that of a cubic
crystal.  The corresponding representations are $E_g$ with
two dimensions and $T_{2g}$ with three
dimensions.\cite{Cheng02}  To illustrate
this, Fig.~\ref{FIG1} shows how the frequencies vary as the
amount of anisotropy is continuously varied.  Let $C_{11a}$,
$C_{12a}$, and $C_{44a}$ be the elastic constants of
directionally averaged gold.  These are based on the
directionally averaged sound speeds of 3330~m/s and 1250~m/s.
In particular, $C_{11a}$ = (3330 m/s)(3330 m/s)(19293 kg/m$^3$)
and $C_{44a}$ = (1250 m/s)(1250 m/s)(19293 kg/m$^3$).
Also, since isotropy is assumed, $C_{12a} = C_{11a} - 2 C_{44a}$.
The elastic constants of true monocrystalline bulk gold are
$C_{11b}$, $C_{12b}$, and $C_{44b}$ as given earlier in this
section.  The frequencies in Fig.~\ref{FIG1} are done with
elastic constants $C_{ij}$ = $x C_{ija} + (1-x) C_{ijb}$.
Note how the isotropic (SPH,2) mode at $\omega R = 3320$ m/s
continuously splits into two modes.

\section{Matrix Effect}

The molecular dynamics calculation results shown in
Tab.~\ref{TAB1} assume a gold sphere with a free surface.
However the gold nanospheres reported in the experimental
results of this paper are embedded in amorphous silica.
As a result, the outer surface of the gold spheres is
not free.  The silica will affect the vibrational
frequencies.

The complex frequency model (CFM)
\cite{Dubrovskiy81,Verma99,Saviot04,Murray04} calculates
the complex-valued vibrational frequencies of a sphere
embedded in an infinite matrix.  Both the sphere and the
matrix are assumed to be continuous, homogeneous and
isotropic.  Since vibrational modes of the gold sphere
are damped by the propagation of sound waves into the
matrix, the modes have complex valued frequencies.
The imaginary part of the frequency is related to the
damping time of the mode.

To treat gold as an isotropic material requires an
approximation to be made about its elastic properties.
We use directionally averaged speeds of sound.  The
longitudinal and transverse speeds are separately
averaged.  The results are 1250 m/s~and 3330~m/s.

Amorphous silica is an isotropic material.  We assume
speeds of sound of 3760~m/s and 5950~m/s and a density
of 2.20~g/cc.

Results of CFM calculations are shown in Tab.~\ref{TAB2}.
FSM calculations based on Lamb's 1882 solution for a free
sphere are also given.

\begin{table}
\caption{\label{TAB2}
FSM and CFM frequencies for a directionally averaged
(elastically isotropic) sphere of gold.  For
the complex CFM frequencies, the
gold sphere is embedded in a silica matrix.
The value given is $\omega R$ where $\omega$ is the
frequency in radians per second and $R$ is
the sphere radius in metres.}
\begin{tabular}{|c|c|c|c|c|c|c|}
\hline
$q$  & $\ell$ & $n$ & FSM $\omega R$ & $Re(\omega R)$ & $Im(\omega R)$ & degeneracy \\
\hline
SPH  & 2 & 0 & 3320 & 4261 & 322 & 5 \\
\hline
SPH  & 0 & 0 & 9789 & 10185 & 542 & 1 \\
\hline
\end{tabular}
\end{table}

We can either model the anisotropic elasticity of gold
using molecular dynamics calculations for a free sphere
or else model the effects of the silica matrix using
CFM for an isotropic sphere.  We have no way to
simultaneously model both anisotropy and matrix effects.

In Tab.~\ref{TAB2}, note that the damping of the modes
is relatively small, based on the low imaginary parts
of the CFM frequencies.  However, the (SPH,$\ell$=2) mode
is signficantly upshifted by the silica matrix.  This
is caused by the action of the matrix to provide a
spring-like reaction force acting at the outer surface
of the gold sphere.\cite{audiprl04}  As a result, the
MD calculations which ignore the matrix effect will
require some correction.

\section{Tandetron accelerator}

Gold ion implanation was done using a 1.7~MV Tandetron
accelerator (Model 4117HC) from High Voltage Engineering
Europa, Amersfoort, the Netherlands.  This accelerator has
the ``tandem'' configuration where negative ions generated
by a cesium sputter gold ion source are first accelerated
from ground to the high voltage terminal and there the
negative gold ions are converted to positive gold ions,
while passing through a stripper canal filled with nitrogen
gas. The same high voltage again accelerates the positive
Au$^{++}$ ions to ground potential.  In this way, Au$^{++}$
ions could be generated with energies up to 5.1~MeV with the
Tandetron.

\section{Sample Preparation}

Using the Tandetron ion accelerator, Au clusters were grown
by direct Au-implantation on fused quartz (amorphous silica)
substrates.  Implantation was performed using 1.8~MeV Au$^{++}$
at 1$\times$10$^{-5}$ Pa with a beam current of $\approx$12~mA
m$^{-2}$ to avoid substantial beam heating in the ion fluence
range of 2$\times$10$^{20}$ -- 1$\times$10$^{21}$ m$^{-2}$.
During implantation the substrate was at room temperature.
The diameter of the ion beam is 1~cm, so that the total
beam current is approximately 1$\mu$A.  At this energy, the
average penetration depth of the Au$^{++}$ in the silica is
445.6~nm before they come to rest, with a straggling
(distribution width) of 55.4~nm.\cite{SRIM13}  The
advantage of using such a high beam energy
is that structure can be created at a much greater depth
below the surface of the silica substrate.  High temperature
air annealing at 873~K for 1 hour was performed to promote
growth of the clusters as gold atoms become mobile in the
silica substrate and are able to conglomerate into bulk gold
metal as opposed to isolated gold atoms.

Our study emphasizes acoustic phonons of the implanted
nanospheres and therefore benefits from a narrow size
distribution.  The 1.8~MeV beam energy is not thought to
play a role apart from controlling the implantation depth
of the ions.  Ion implantation often results in wide size
distributions.\cite{Miotello01,Bandourko03}  Implantation
temperature\cite{Strobel99} and post-implantation
annealing\cite{Fukumi94} are known to play an important
role in controlling average particle size and size
distribution.  The current density of the ion beam affects
average size and size distribution when implanting silica
with Ag\cite{Dubiel03} and Cu\cite{Magruder94} and is
expected from simulations\cite{Strobel99} to play a similar
role for gold.  Laser pulse annealing\cite{Wood93} is also
known to be able to narrow the size distribution.

\section{Sample Characterization}

\begin{figure} 
\includegraphics[width=\columnwidth]{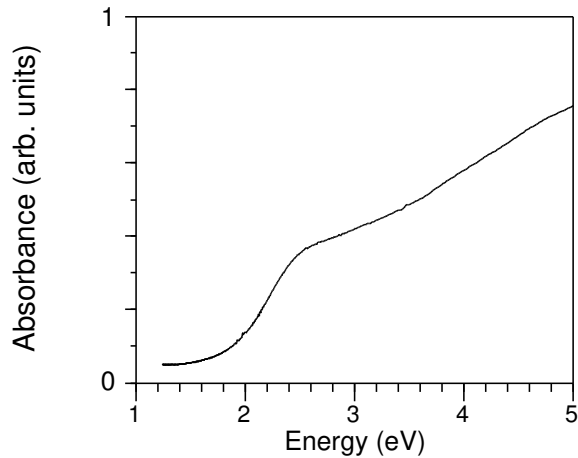}
\caption{\label{FIG2}
Light absorbance as a function of photon energy is shown
for post-annealed Au nanoclusters in silica matrix grown
at fluence 1$\times$10$^{21}$ m$^{-2}$.  Structure around
2.4~eV is attributed to the dipolar surface plasmon
resonance of a gold sphere embedded in silica.}
\end{figure}

Optical absorption spectra were recorded at room temperature
in the range of 200 -- 1100~nm using a Hewlett-Packard
diode-array UV-visible (UV-Vis) spectrophotometer (Model
8453) with necessary signal correction for the substrate.
Figure~\ref{FIG2} shows broad peak structure corresponding
to the frequency of the dipolar surface plasmon resonance of
a gold sphere in a silica matrix.  This is evidence of the
shape of the nanoparticles being approximately spherical.
The surface plasmon resonance of an ellipsoidal sphere would
split into higher and lower frequencies.

\begin{figure} 
\includegraphics[width=\columnwidth]{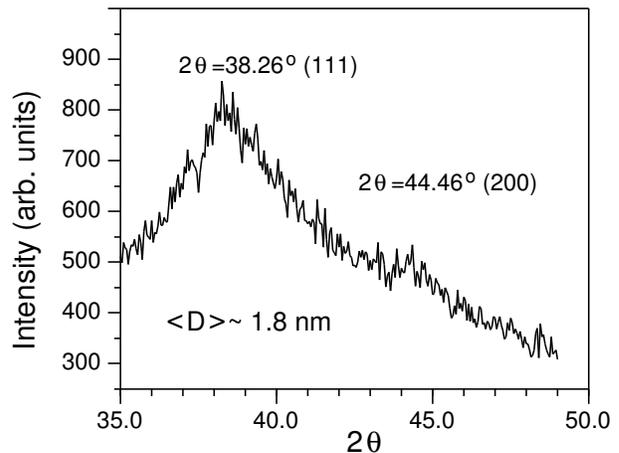}
\caption{\label{FIG3} 
Grazing incidence X-ray diffraction data is shown for
gold nanospheres embedded in silica, using copper
alpha X-rays.  Two expected peak positions are indicated
based on the fcc structure and 0.40786~nm lattice constant
of bulk gold.  The indicated averaged diameter estimated is
based on Scherrer's formula applied to the (111) peak.}
\end{figure}

Grazing incident X-ray diffraction (GIXRD) was done using Cu
K$\alpha$ X-rays (wavelength = 0.15406~nm), as shown in
Fig.~\ref{FIG3}.  Both the (111) and (200) peaks were
observed, with (111) being the stronger feature.  The
location of the (111) peak closely matches that expected
based on a fcc unit cell size of 0.40786~nm for gold at room
temperature.  This was done in 0.05~degree steps with 25~s
accumulation time per data point.  This is an
$\Omega$-2$\theta$ scan. $\Omega$ is 0.8~degrees.  In other
words, the incident X-rays enter the sample at 89.2~degrees
from normal incidence.

X-ray diffraction is an indication of the crystalline
domain size rather than the size of the nanosphere.
However since annealing at 873~K was done, it is expected
that the nanosphere would consist of a single crystal
domain, and the size from GIXRD is interpreted as the
size of the nanosphere.

\begin{figure} 
\includegraphics[width=\columnwidth]{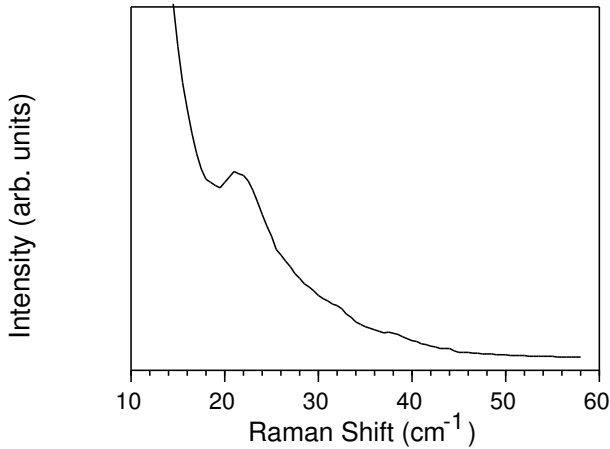}
\caption{\label{FIG4} 
Room temperature low-frequency Raman spectra for
post-annealed Au nanoclusters in a silica matrix.
The 4~cm$^{-1}$ wide peak at 22~cm$^{-1}$ is
attributed to thermal spheroidal quadrupolar vibrations
of the gold sphere.  The broadening of the peak may be
primarily due to damping of the vibration due to
mechanical contact with the silica matrix.}
\end{figure}

Low-frequency Raman scattering studies were done at room
temperature in the back scattering geometry, using
vertically polarized 488~nm line of an argon ion laser
(Coherent, USA) with 200~mW power.  Unpolarized scattered
light from the sample was dispersed using a double
monochromator (Spex, model 14018) with instrument resolution
of 1.4~cm$^{-1}$ and detected using a cooled photomultiplier
tube (FW ITT 130) operated in the photon counting mode.
See Fig.~\ref{FIG4}.

A strong and relatively narrow Raman peak is seen in
Fig.~\ref{FIG4} with its peak at 22~cm$^{-1}$.  The full
width at half maximum (FWHM) of this peak is approximately
4~cm$^{-1}$.  This is relatively narrow compared to Raman
peaks seen in other systems of embedded nanoparticles.
The center of the Raman peak corresponds to an angular
frequency $\omega$ = 4.15$\times$10$^{12}$~rad/s.

\section{Discussion}

The low frequency Raman spectrum of our sample in
Fig.~\ref{FIG4} showed a prominent peak with an angular
frequency of 4.15$\times$10$^{12}$ rad/s.  The average
radius of the particles, as estimated from GIXRD in
Fig.~\ref{FIG3} using Scherrer's formula was 0.90~nm.  The
resulting experimental value of $\omega R$ is 3732~m/s.

The existence of a vibrational mode does not mean that it
will appear in a Raman spectrum.  For an isotropic sphere,
only the (SPH,0) and (SPH,2) modes can be Raman active based
on group theoretical considerations\cite{Duval92} and in
practice the intensity of the (SPH,0) is much smaller.  More
detailed analysis, not presented here, is necessary to
predict the relative Raman activity of the two different
frequencies into which the (SPH,2) mode has been split by
elastic anisotropy.\cite{Bachelier04,Cheng02}

It is expected that one of the two modes with $\omega R$ =
2342~m/s and 3779~m/s as shown in Tab.~\ref{TAB1} will be
the dominant feature in the Raman spectrum, but these values
from MD omit the matrix effect.  To estimate the effect of
the matrix, we rely on comparison of FSM and CFM results in
Tab.~\ref{TAB2}, which shows that the matrix effect
increases the frequency of the (SPH,$\ell$=2) mode by a
factor of 1.28.  Applying this to the MD results, the two
expected locations of Raman peaks are $\omega R$ = 3006~m/s
and 4837~m/s.  One is 20\% lower than the location of the
experimental Raman peak while the other is 30\% higher.

Based on the cubic symmetry, for modes with $T_{2g}$
symmetry, the Raman tensor is off-diagonal while for
modes with $E_g$ symmetry, the Raman tensor is diagonal.\cite{Cheng02}
In other words, $T_{2g}$ modes will only contribute to the
depolarized (VH) portion of the Raman spectrum while
$E_{g}$ modes only contribute to the polarized (VV) part.
The Raman spectrum in our measurements is polarized.
Thus, $\omega R$ is experimentally 3732~m/s and
theoretically 4837~m/s.

A possible resolution of this poor agreement is that the
gold nanoparticles are loosely attached to the silica
matrix.  In that case, FSM frequencies from Tab.~\ref{TAB1}
would be appropriate.  Since the $E_g$ mode frequency
is 3779, the agreement with the experimental value of
3732~m/s is compelling.  Furthermore, the next paragraphs
show an independent reason to believe this.

CFM calculations of the width of the peak show that it
should be quite narrow.  Based on the CFM value of
$\omega R$ of the (SPH,2) mode given in Tab.~\ref{TAB2}, the
expected FWHM width of a Raman peak at 22~cm$^{-1}$ would be
3.3~cm$^{-1}$, which is only slightly smaller than the
observed FWHM width in Fig.~\ref{FIG4} of 4~cm$^{-1}$.
The increased peak width can be attributed to size variation
of the gold nanoparticles within the sample as well as
instrumental broadening.  It is plausible to assume that the
Raman width due to particle size variation, the Raman width
due to mode damping, and the Raman instrumental width
(1.4~cm$^{-1}$) combine squarewise to give the width of
the observed Raman spectrum.  In that case, the Raman peak
FWHM width due to particle size variation is just 1.8~cm$^{-1}$,
since $4^2 - 3.3^2 - 1.4^2 = 1.8^2$.  We assume the
distribution of particle radius is Gaussian with one
standard deviation $\sigma$ and mean $R_{avg}$.  The FWHM of
a Gaussian distribution is $2\sqrt{2\,ln 2} \, \sigma$.
Thus, the rms variation of particle radius is estimated
to be $\sigma = 0.034 R_{avg}$. 

It is difficult to believe that the particle radius
variation in our sample is really so small, particularly
since ion implantation experiments have never given such a
tight size distribution.  An alternative is that CFM
overestimates the true broadening due to damping.  If the
interface between the gold and the silica is not a tight
bond then the broadening effect of the silica matrix is
reduced based on a model of the interface as a thin soft
layer.\cite{audiprl04}  Since gold does not readily form
chemical bonds with silicon or oxygen the adhesion between
the gold and silica is likely of Van der Waals type.

Acknowledgements: The authors would like to thank 
P.~Ramamurthy of the National Centre for Ultrafast
Processes, University of Madras, India for his co-operation
in UV-Vis study.  We thank V.~S.~Sastry and B.~K.~Panigrahi
of MSD, IGCAR for their contributions in structural and
implantation studies, respectively.  We also thank
B.~Viswanathan of MSD, IGCAR for his encouragement in
pursuing this work.

\end{document}